# Pressure-Induced Chemical Bonding Effects on Lattice and Magnetic Instabilities in Antiferromagnetic Insulating $CaMn_2Sb_2$

Matt Boswell[1], Antonio M. dos Santos[2], Mingyu Xu[1], Madalynn Marshall[3], Su-Yang Xu[4], Weiwei Xie[1*]

1. Department of Chemistry, Michigan State University, East Lansing, MI 48824 USA
2. Neutron Scattering Division, Oak Ridge National Laboratory, Oak Ridge, TN 37831 USA
3. Department of Chemistry and Biochemistry, Kennesaw State University, Kennesaw, GA 30144 USA
4. Department of Chemistry and Chemical Biology, Harvard University, Cambridge, MA 02138, USA

Corresponding Author: Weiwei Xie (xieweiwe@msu.edu)

*Abstract*

Exotic quantum phenomena often emerge near an electronic delocalization transition (EDT) from an antiferromagnetic insulating phase to a strongly correlated metallic state under pressure. We report the pressure-induced structural and magnetic evolution of the antiferromagnetic insulator $CaMn_2Sb_2$. Single-crystal X-ray diffraction reveals a first-order phase transition near 5.4 GPa from a trigonal *P*-3*m*1 structure to a monoclinic *P*2$_1$/*m* phase, accompanied by a ~7% volume collapse. Residual electron density analysis at intermediate pressures reveals charge localization along Mn-Sb chains, signaling electronic instability preceding the structural transition. Bonding analysis indicates anisotropic Mn-Sb orbital reconfiguration under pressure, driving a distorted square-pyramidal geometry. Neutron scattering confirms the transition and identifies a pressure-induced incommensurate magnetic order, distinct from the ambient antiferromagnetic state. In the monoclinic phase, zigzag Mn chains exhibit antiferromagnetic coupling along the *ac*-plane, enabled by enhanced orbital overlap. These results establish $CaMn_2Sb_2$ as a model system for studying the coupling of structural distortion, charge redistribution, and magnetic order in layered Mn pnictides under pressure.

**Keywords**

# Introduction

Exotic quantum phenomena, such as superconductivity and charge density waves (CDWs), often emerge near an electronic delocalization transition (EDT) from an antiferromagnetic insulating phase to a strongly correlated metallic state.[1,2] A well-known example lies in the iron-based high $T_c$ superconductors, where the parent compounds typically reside just on the metallic side of a Mott-like EDT.[3,4] This proximity has inspired the hypothesis that higher superconducting transition temperatures ($T_c$) might be realized if a new Fe-based parent compound could be synthesized on the insulating side of the EDT.[5] Given the hereto unsuccessful quest to enhance $T_c$ within the Fe-based systems, attention has shifted toward the structurally-related layered manganese pnictides.[6,7] Unlike their Fe analogs, the parent Mn-based compounds are already predominantly antiferromagnetic insulators, making them promising candidates for tuning across EDTs via chemical doping or external pressure. However, despite considerable efforts, inducing superconductivity in these systems has remained elusive. For instance, LaMnPO undergoes metallization under pressures of ~10 GPa.[8] $BaMn_2As_2$ likewise, exhibits pressure- or doping-induced metallization, yet superconductivity has not been observed in any of these cases.[9]

This raises key questions: How do Mn-based antiferromagnetic insulators respond to electronic instability under doping or pressure? And why do such systems tend to develop alternate quantum states instead of superconductivity?

To address these questions, we investigate $CaMn_2Sb_2$, an antiferromagnetic insulator crystallizing in a layered trigonal structure.[10,11] At ambient conditions, $CaMn_2Sb_2$ exhibits a direct optical band gap of ~1 eV, with a smaller thermal activation gap.[12] Above the Néel temperature ($T_N$ ~ 85 K), long-range antiferromagnetic order sets in, followed by the appearance of a weak ferromagnetic phase around 210 K.[13, 14, 15] Structurally, $CaMn_2Sb_2$ belongs to the broader $AB_2X_2$ family, which includes the widely studied tetragonal $ThCr_2Si_2$-type structure, adopted in many iron-based superconductors like $BaFe_2As_2$.[16,17] In this family, the A-site is typically occupied by an alkali, alkaline earth, or rare-earth element, the B-site by a transition metal or a main group element, and X is generally a Group 15, 14, or occasionally Group 13 element. In addition to the tetragonal variant, a trigonal 122-type structure also exists within the $AB_2X_2$ family. This structure arises from the distortion of the T-Pn (T = transition metal, Pn = P, As, Sb, Bi) coordination environment through either the elongation or contraction of one T-Pn bond. Notably, this trigonal variant has been observed in several dozen compounds, establishing it as another prominent structural prototype in this class of materials. This trigonal structure appears to preferentially form when the B-site element possesses a $d^0$, $d^5$, or $d^{10}$ electronic configuration. For instance, in topological Nodal-line semimetal $Mg_3Bi_2$ ($MgMg_2Bi_2$), the formal oxidation states $Mg^{2+}(Mg^{2+})_2(Bi^{3-})_2$, yield a $d^0$ configuration for $Mg^{2+}$.[18] Similarly, $Ca^{2+}(Mn^{2+})_2(Sb^{3-})_2$ and $Sr^{2+}(Zn^{2+})_2(As^{3-})_2$ exemplify the $d^5$ and $d^{10}$ configurations, respectively.[10] Notably, the absence of short X-X contacts, which means no polyanionic cluster forms among the main group elements in the trigonal 122 structure justifies treating Bi, Sb and As as $X^{3-}$ anions, respectively.[19]

From a chemical perspective, pressure alters interatomic bonding, potentially inducing structural phase transitions distinct from those anticipated purely from an electronic standpoint.[20] To probe the electronic and magnetic response of $CaMn_2Sb_2$ under pressure, we performed high-pressure single-crystal X-ray diffraction and neutron (powder) scattering measurements. Our high-pressure single-crystal X-ray diffraction experiments reveal the structural evolution of $CaMn_2Sb_2$ under compression, while neutron scattering measurements up to 5.9 GPa and down to 85 K uncover the emergence of one-dimensional incommensurate magnetic ordering. Notably, rather than superconductivity, the system develops electronic instabilities and modulated magnetic states-highlighting the richness of quantum phases accessible in Mn-based layered pnictides under extreme conditions.

## Experimental Details

**High Pressure Single Crystal X-ray Diffraction:** Single crystals of $CaMn_2Sb_2$ were synthesized using a Sn flux method following a previously reported procedure. The crystal structure was confirmed by room-temperature X-ray diffraction under ambient pressure. Magnetic ordering was validated through resistivity measurements (**Fig. S1**), which reveal a magnetic transition near 85 K, in agreement with earlier reports. High-pressure single-crystal X-ray diffraction measurements were performed up to 6 GPa using a Rigaku XtaLAB Synergy-S diffractometer equipped with a Mo Kα radiation source ($\lambda_{K\alpha}$ = 0.71073 Å) with the beam size 100um × 100um. A high-quality single crystal of $CaMn_2Sb_2$ (~20um × 20 um) was first mounted on a nylon loop using Paratone oil and measured at ambient pressure to confirm the crystal structure and orientation. The sample was then loaded into a Diacell One20DAC (Almax-easyLab) equipped with extra-wide aperture seats (120 °) with diamond anvils featuring 500 μm culets. A stainless-steel gasket (250 μm thick) was pre-indented to a final thickness of ~75 μm, and a ~200 μm diameter hole was drilled using an electric discharge machining (EDM) system to form the sample chamber. A 4:1 methanol-ethanol mixture was used as the pressure-transmitting medium to ensure hydrostatic conditions.[21] Pressure inside the cell was calibrated using the R1 fluorescence line of a ruby sphere placed adjacent to the crystal.[22] To confirm reproducibility and reliability of the results, the high-pressure diffraction experiment was performed twice under identical conditions.

**High Pressure Neutron Diffraction:** High-pressure neutron powder diffraction experiments were performed at the Spallation Neutron Source (SNS) at Oak Ridge National Laboratory (ORNL) on the SNAP beamline. Approximately 500 mg of finely ground $CaMn_2Sb_2$ powder was loaded into a single-toroidal anvil Paris-Edinburgh press equipped with a liquid nitrogen cooling system. The sample was loaded without a pressure-transmitting medium to avoid ambiguity on the chemical stability and excess signal at low temperature from frozen media precluding tracking of the magnetic structure. Neutron scattering data were collected using a wavelength band about 3.5 Å wide and centered at 2.1 Å. The two detector banks were positioned at 90° and 50° to capture both nuclear and magnetic diffraction signals, respectively. Pressure was incrementally increased in steps of 75 bar (≈ 0.5 GPa) up to a maximum of 1050 bar (≈ 5.9 GPa), with temperature-dependent measurements collected at selected pressure points. Pressure calibration was carried out using the equation of state for $CaMn_2Sb_2$ obtained from independent single-crystal high-pressure X-ray diffraction measurements. Nuclear diffraction patterns from the 90° detector bank were refined using the GSAS-II software package[23], while magnetic diffraction data from the 50° bank were analyzed using the FullProf suite.[24] Symmetry analysis, including the identification of magnetic space groups and irreducible representations, was conducted using the SARAh software.[25]

## Results and Discussion

**High Pressure X-ray diffraction:** At ambient pressure, $CaMn_2Sb_2$ crystallizes in a layered trigonal structure with space group $P\text{-}3m1$ (Space Group, No. 164). The structure consists of alternating layers of $Ca^{2+}$ and $[Mn_2Sb_2]^{2-}$, in which Mn atoms form a buckled honeycomb lattice with an intralayer Mn-Mn distance of approximately 3.202 Å. Upon compression, single-crystal x-ray diffraction measurements reveal that $CaMn_2Sb_2$ retains the $P\text{-}3m1$ symmetry up to at least 4.5 GPa. However, at 5.4 GPa, a first-order displacive phase transition to a monoclinic structure with space group $P2_1/m$ is observed. The complete crystallographic parameters at various pressures are summarized in **Tables S1** and **S2**. To further characterize the pressure response, the volume per formula unit was plotted as a function of pressure and fitted using the second-order Birch-Murnaghan equation of state, as shown in **Fig. 1a**. Below 5.4 GPa, the material remains in the $P\text{-}3m1$ phase, with the equation of state fitting yielding $V_0 = 133.6 \pm 0.2$ Å$^3$ and $B_0 = 42 \pm 2$ GPa, which is similar with the ones in isoelectronic $CaMn_2Bi_2$.[26] A significant volume collapse (~7%) is observed at 5.4 GPa, accompanied by a marked reduction in crystallinity, which persists up to 7.4 GPa, the highest pressure reached. The increase in GOF and R factors with increasing pressure is a systematic and unavoidable feature of high-pressure single-crystal X-ray diffraction experiments performed in diamond anvil cells, particularly for compounds containing heavy elements such as Sb. Because we can't perform numerical absorption corrections to the crystal in DAC. **Figs. 1b** and **1c** compare the low-pressure and high-pressure crystal structures of $CaMn_2Sb_2$. In the ambient-pressure phase (**Fig. 1b**), Mn atoms form a buckled honeycomb lattice, with each Mn coordinated to four Sb atoms in a square-pyramidal geometry, with the Mn atom occupying the apex position of the pyramid. Each Sb atom bonds to four Mn atoms, with three Mn-Sb bond lengths 2.782 Å and one slightly longer at 2.784 Å. Under high pressure shown in **Fig. 1c**, the structure undergoes a symmetry-breaking transition to a monoclinic phase, resulting in two unique Mn sites and two unique Sb sites, compared to only one Mn and one Sb site at ambient pressure. Furthermore, the local atomic environments are significantly altered. One of the Sb atoms forms one Mn-Sb bond, while the other adopts a distorted square-pyramidal coordination environment, bonded to one Sb atom on the apex position and four surrounding Mn atoms. In this distorted geometry, two Mn-Sb bond lengths are shortened to approximately 2.64 Å, while the other two are extended to about 2.85 Å. This reflects a clear deviation from the nearly regular square pyramidal coordination at ambient pressure.

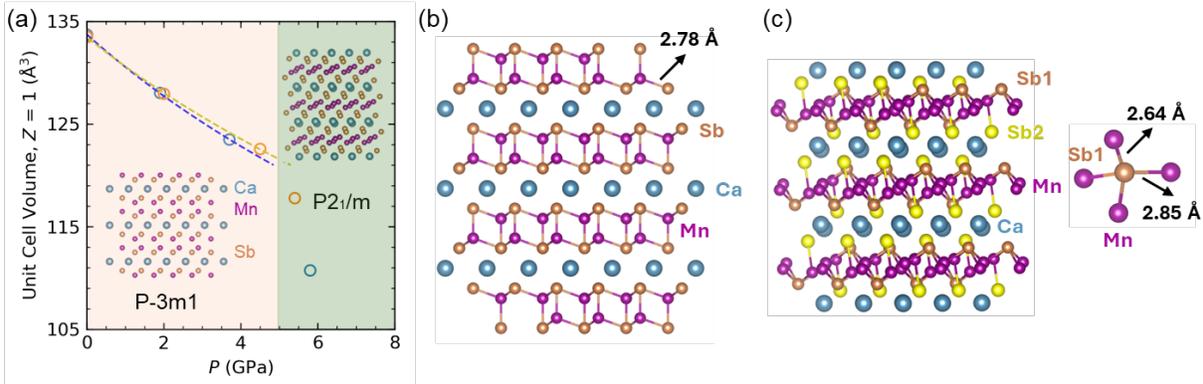

**Fig. 1** (*a*) Pressure dependence of the volume per formula unit of $CaMn_2Sb_2$ fitted using second-order Birch-Murnaghan equation of state. (*b*) Crystal structure of $CaMn_2Sb_2$ at ambient pressure (*c*) Crystal structure of $CaMn_2Sb_2$ at 5.9 GPa, with Sb-Mn bond distances indicated.

**Residual Electron Density analysis:** To gain deeper insight into the pressure-induced evolution of the electronic structure in $CaMn_2Sb_2$, residual electron density maps were analyzed at 4.5 GPa, a pressure at which the compound remains in the trigonal phase but lies near the structural phase transition boundary. **Figs. 2*a*** and **2*b*** present projections of the residual electron density along the *c*-axis and within the *ab*-plane, respectively. When viewed down the *c*-axis (**Fig. 2*a***), pronounced residual electron density is observed at the Mn and Sb atomic positions within the buckled honeycomb lattice, while negligible residual density is detected at the Ca sites. When viewed along the *ab*-plane (**Fig. 2*b***), the Mn-Sb layers exhibit strong residual electron density forming two well-defined linear chain motifs, spatially separated by intervening Ca layers. These observations indicate an anisotropic charge distribution and suggest the development of linear electronic features within the Mn-Sb sublattice, pointing to emerging charge instabilities as the system approaches the phase transition under pressure.

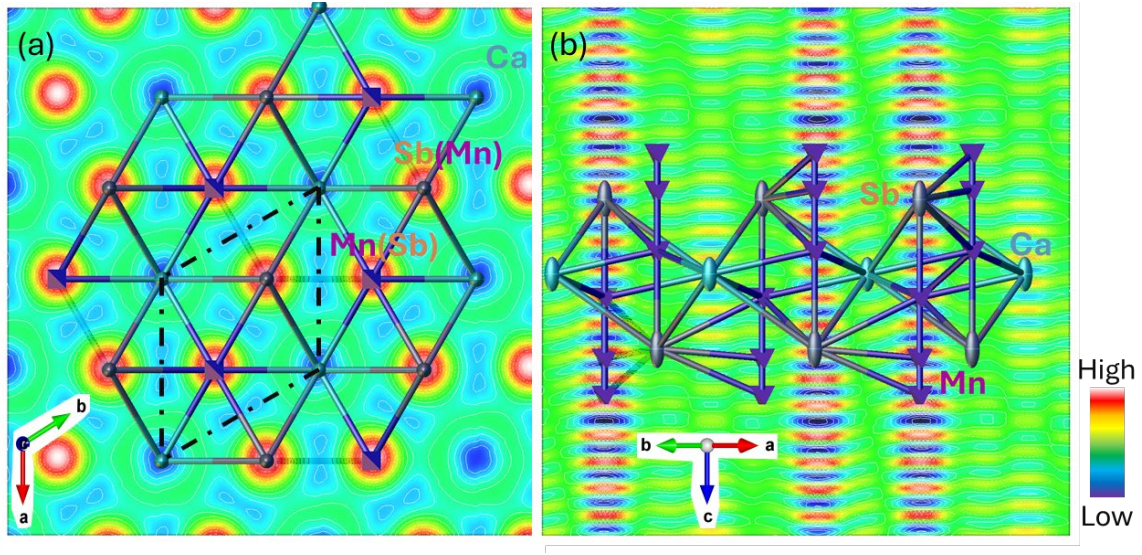

**Fig. 2.** Residual electron density map of CaMn$_2$Sb$_2$ viewed **(a)** along the *c*-axis and **(b)** within the *ab*-plane.

**Bonding Analysis and Structural Stability under Pressure:** Hoffmann and Chong have provided a theoretical explanation for the apparent preference for $d^0$, $d^5$, or $d^{10}$ configurations at the B-site in trigonal type compounds.[16] To further elucidate the bonding interactions and orbital characteristics in CaMn$_2$Sb$_2$ and potential pressure effects, we analyze both its experimentally observed trigonal structure and a hypothetical tetragonal analogue. We begin with the tetragonal 122-type structure, in which each B-site atom (e.g., Mn) is tetrahedrally coordinated by four X atoms (e.g., Sb) and also has four nearest-neighbor B atoms. The X-site atoms are likewise four-coordinate but adopt a less common square-pyramidal coordination geometry. In contrast, the trigonal 122-type structure features a distorted B-X coordination environment, where the four B-X bond lengths are no longer equivalent. This geometry resembles an "umbrella" motif: three shorter B-X bonds form the "ribs" of the umbrella, while the fourth, axial bond, represents the "shaft". The degree of distortion in the trigonal structure appears to be correlated with the electronic configuration of the B-site atom. In compounds where B has a $d^0$ or $d^{10}$ configuration (e.g., Mg$^{2+}$ or Zn$^{2+}$), the axial "shaft" bond is significantly longer than the equatorial "rib" bonds. However, for $d^5$ systems such as those containing Mn$^{2+}$, the bond length differences are much less pronounced. In some cases, the axial bond is even shorter than the equatorial ones, suggesting a different bonding interaction that reflects the partially filled *d*-shell.

A schematic molecular orbital diagram for the SbMn$_4$ tetrahedron is shown in **Fig. S3a**. The four lowest-energy orbitals are primarily derived from Sb *s* and *p* orbitals, which hybridize with symmetry-adapted Mn local radial σ orbitals (notably involving $d_{z^2}$, *s*, and $p_z$ components). Above these levels lies the Mn *d* orbital manifold. At the bottom of this *d*-block are orbitals with predominantly local Mn σ and π character, which are largely nonbonding with respect to Mn-Sb interactions. The upper part of the *d* block contains antibonding counterparts to the four lowest Mn-Sb bonding orbitals, with contributions mainly localized on Sb. However, the most significant Mn-Sb antibonding interactions originate from higher-energy Mn 4*s* and 4*p* orbitals. Transformation of the Mn-Sb bonding environment from a tetrahedral to a trigonal geometry, as illustrated in **Fig. S3b**, alters the orbital degeneracy. In this distorted geometry, the Sb *s*, $p_x$, and $p_y$ orbitals largely preserve their overlap with Mn orbitals and remain energetically stable. In contrast, the $p_z$ orbital undergoes increased antibonding interaction with the Mn "rib" orbitals, leading to an upward shift in energy – effectively elevating this orbital to the top of the Mn *d* block. A lower-lying orbital, originally of local Mn character, correspondingly shifts downward in energy to replace it. The abundance of energetically similar Mn *d* orbitals imparts a degree of topological flexibility to the SbMn$_4$ motif, allowing the bonding framework to accommodate structural distortions with minimal energetic penalty.

These features are further manifested under applied pressure. Experimentally, the three Mn-Sb "rib" bond distances show minimal compression (from 2.783 Å at ambient pressure to 2.781 Å at 4.5 GPa), whereas the axial "shaft" Mn-Sb bond shortens significantly (from 2.784 Å to 2.536 Å). This anisotropic compression implies an enhanced bonding interaction between Mn $d_{z^2}$ and Sb $p_z$ orbitals along the axial direction. Concurrently, the three antibonding σ* interactions involving Mn $d_{z^2}$ and Sb $p_z$ are distorted, suggesting a reconfiguration of bonding topology. Meanwhile, the comparatively weak in-plane *p* bonding interactions are further destabilized. This evolution ultimately leads to a distorted SbM$_4$ square-pyramidal configuration and linear Sb-Mn bonding, consistent with the monoclinic structural phase observed in CaMn$_2$Sb$_2$ near 5.9 GPa.

**High-Pressure Neutron Diffraction**: Neutron diffraction measurements were performed to investigate the pressure-induced structural phase transition in CaMn$_2$Sb$_2$, and its possible magnetic ordering. As pressure increases from ambient conditions to approximately 3 GPa, only the low-pressure trigonal phase is observed, in agreement with both x-ray diffraction data and equation-of-state fits. Rather than directly comparing transition pressures, we note that the refined unit-cell volumes obtained from neutron diffraction fall within the same volume range as those measured by X-ray diffraction prior to the phase transition. This correspondence suggests that the structural stability regime of the trigonal phase is comparable between the two methods. We note that the neutron measurements were conducted without a pressure-transmitting medium and therefore may involve non-hydrostatic stress; however, the consistency in observed volumes indicates that such effects do not significantly shift the phase boundary within the investigated range. Around 4 GPa the monoclinic phase begins to emerge within the powder diffraction pattern, marking the onset of the structural transition. Between ~4 GPa and ~5 GPa, both the ambient trigonal phase and the high-pressure monoclinic phase coexist, with the latter becoming progressively dominant. Above ~5 GPa, only the monoclinic phase remains detectable. This phase coexistence is clearly visible in **Fig. 3**. The presence of both phases may be attributed to deviatory stresses across the sample; a phenomenon also observed in the corresponding single-crystal X-ray experiments. Another contributing factor could be the substantial volume collapse associated with the phase transition, which likely imposes a significant kinetic barrier to complete the structural transformation.

In the high-pressure regime, the *hkl* reflection near 3.2 Å becomes substantially broadened, complicating structural refinements. It remains unclear whether this broadening arises from low crystallinity of the sample, pressure inhomogeneity, slow kinetics of the phase transition, or intrinsic disorder within the high-pressure phase. Notably, full convergence of structural fits to this broad peak using the higher-resolution 90° detector bank proved challenging. However, magnetic peak fitting using data from the lower-resolution 50° detector bank was more robust and did not suffer from this limitation. Pressure estimates derived from both internal lead calibrant, and internal equation-of-state determination are in reasonable agreement, supporting the reliability of the pressure calibration and phase assignment.

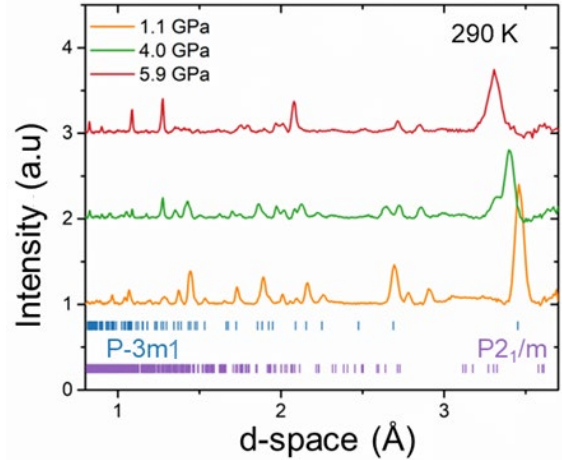

**Fig. 3.** Neutron powder diffraction patterns of CaMn$_2$Sb$_2$ collected at 290 K under applied pressures of 1.1 GPa, 4.0 GPa, and 5.9 GPa, highlighting structural evolution *vs* pressure.

**Pressure induced Magnetic Ordering in CaMn$_2$Sb$_2$:** At low pressures, magnetic reflections in CaMn$_2$Sb$_2$ are observed around 85 K, consistent with previously reported neutron scattering and magnetic susceptibility measurements.[12] At ambient pressure CaMn$_2$Sb$_2$ lies near a magnetic triple critical point between two different spiral orders and an antiferromagnetic order, which is the ground state.[14] The antiferromagnetic order that is stabilized is commensurate with propagation vectors of (0, 0, 0) aligning with spins confined into the ab-plane.[11,12] Lying near the triple critical point modifying the interaction ratio between nearest-neighbor and further neighbors can result in the spiral orders being stabilized. However, up to 1 GPa the magnetic structure remains within the antiferromagnetic order. The persistence of Néel-type order upon increased pressure is reminiscent of behavior observed in the AYbSe$_2$ (A = Na, K, Cs) family of materials. Upon increasing pressure, the nearest neighbor interaction would strengthen considerably more than further neighbors resulting in the exchange ratio diminishing.[27,28] In CaMn$_2$Sb$_2$ lowering of the exchange ratio would push the system further away from the triple critical point and deeper into the antiferromagnetic order. These results suggest that pressure can be used in a control way to diminish exchange interaction ratios to reach new magnetic states.

After the structural transition new magnetic peaks emerge, corresponding to an incommensurate magnetic state that persists to higher temperatures compared to the antiferromagnetic order. Utilizing a k-search led to propagation vectors of (0.8, 0.33, 0.2), however, allowing free refinement found best fit propagation vectors of (0.85, 0.32, 0.2). In contrast to isostructural compound $CaMn_2Bi_2$, which exhibits incommensurate propagation vectors of (0. 0.48, 0.13), highlighting key differences in magnetic dimensionality and coupling.[29] In comparison $Ba_2FeSbS_5$ was found to contain an incommensurate magnetic lattice that was not observed in the isostructural $Ba_2FeBiS_5$, believed to be induced by the smaller bond distance and stronger exchange between further neighbors causing a higher degree of magnetic frustration.[30]

At ambient pressure, $CaMn_2Sb_2$ orders magnetically near 90 K, while $CaMn_2Bi_2$ orders around 150 K; however, under applied pressure their magnetic ordering temperatures begin to converge.[10,11] A similar trend, where Bi-containing analogues exhibit higher ordering temperatures than their Sb counterparts, has been reported in other ternary transition-metal pnictides.[30,31,32] This behavior is commonly attributed to the larger radial extent and enhanced covalency of Bi 6p orbitals, which strengthen Mn-Bi-Mn superexchange interactions relative to Mn-Sb-Mn pathways. Additionally, the stereochemically active $6s^2$ lone pair on Bi can influence local bonding geometry and hybridization, thereby indirectly modifying magnetic exchange pathways.[33,34] In $CaMn_2Sb_2$, the comparatively weaker covalency of Sb 5p orbitals and stronger competition between nearest- and further-neighbor interactions may enhance magnetic frustration, leading to a lower magnetic ordering temperature. As pressure is applied and interatomic distances decrease, covalent interactions in both compounds are enhanced, reducing the disparity between Sb- and Bi-mediated exchange. Pressure-induced modifications of bond angles and orbital hybridization further reorganize exchange pathways, resulting in similar effective magnetic energy scales and, consequently, comparable ordering temperatures.[35] Although the inert pair effect in Bi is primarily structural rather than directly magnetic in origin, pressure-enhanced hybridization can indirectly influence magnetic exchange through lattice distortions and modified bonding topology.[36]

As temperature increases above the magnetic ordering temperature, the magnetic Bragg peak intensities decrease progressively. In particular, between 90 K and 125 K, the magnetic reflection at 5.25 Å is strongly suppressed, approaching the intensity level of the nuclear reflections. Upon further heating, the magnetic peaks at 4.5 Å and 5.25 Å disappear entirely. As shown in **Fig. 4a**, by 180 K most magnetic reflections are indistinguishable from nuclear peaks, and by 290 K only a weak residual feature near 4 Å remains, which continues to diminish with increasing temperature. At larger d-spacings, the resolution decreases due to limited detector coverage and the reduced sample volume under compression. Consequently, weak magnetic features in this region may be partially obscured in the powder diffraction pattern.

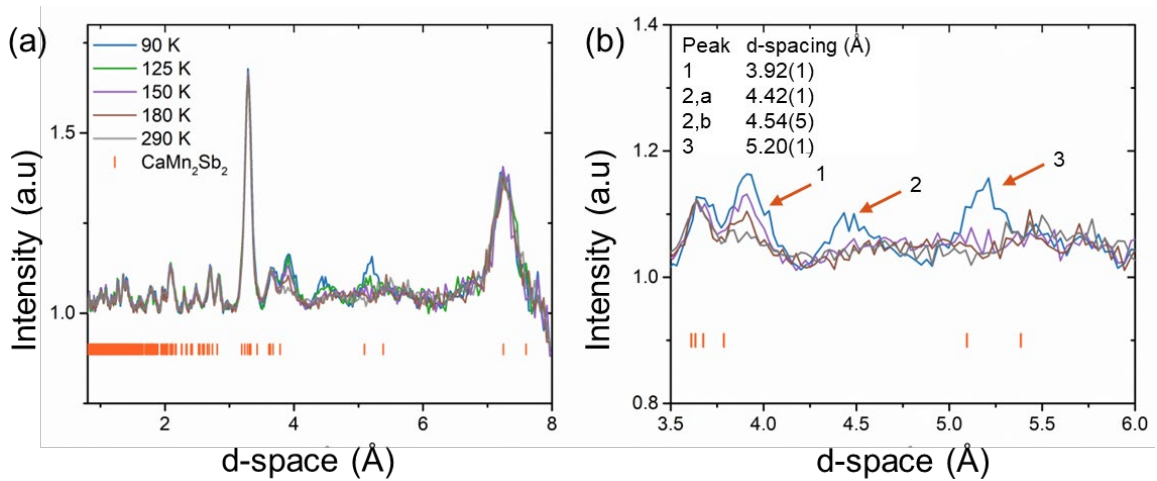

**Fig. 4** (*a*) Temperature-dependent neutron diffraction patterns of CaMn$_2$Sb$_2$ at 5.9 GPa, corresponding to the high-pressure phase. Orange tick marks indicate the nuclear Bragg peaks. (*b*) Enlarged view of the diffraction patterns in the *d*-spacing of 3.5-6 Å region, emphasizing the evolution of magnetic reflections with temperature. Orange arrows point to the magnetic regions, note the peak at 4.5 when fit to a gaussian function denotes two magnetic peaks. The fitted magnetic peaks to a Gaussian function with their corresponding *d*-spacings are provided.

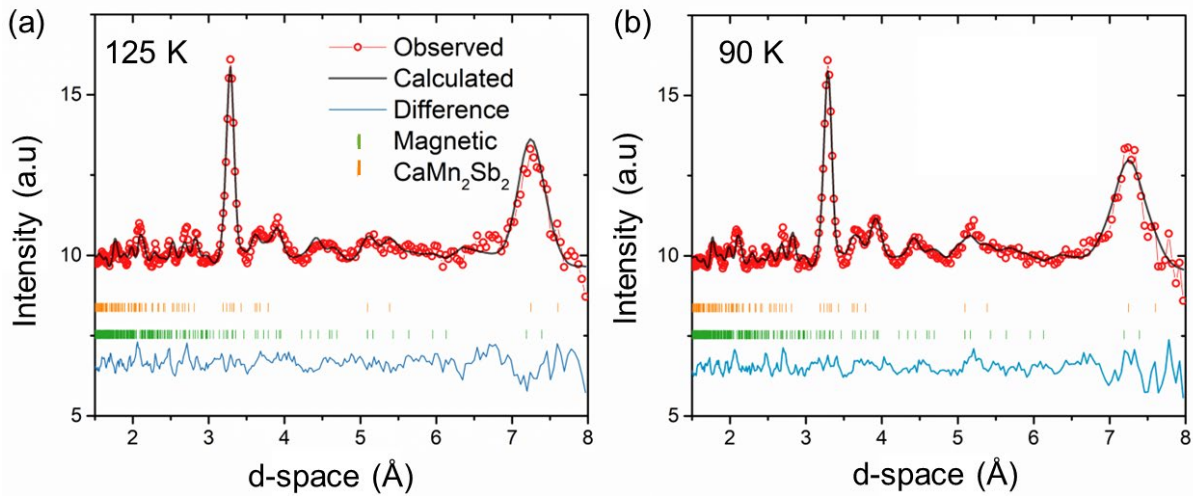

**Fig. 5.** Rietveld refinement of neutron diffraction data for CaMn$_2$Sb$_2$ at 5.9 GPa. (*a*) Combined nuclear and magnetic peak fitting at 125 K. (*b*) Refined pattern at 90 K, illustrating the development of magnetic order at lower temperatures.

Representational analysis based on the refined propagation vector in $CaMn_2Sb_2$ identifies one irreducible representation with three basis vectors (**Table S3**). The monoclinic high-pressure structure contains two inequivalent Mn sites (Mn1 and Mn2), each split into two magnetically distinct sublattices: Mn1$a$ = ($x$, 0.25, $z$), Mn1$b$ = (1-$x$, 0.75, 1-$z$), Mn2$a$ = ($u$, 0.25, $w$), and Mn2$b$ = (1-$u$, 0.75, 1-$w$). Free refinement of the mixing coefficients using the FullProf suite yields a sinusoidally modulated magnetic structure, similar in form to that observed in $CaMn_2Bi_2$ and other Mn-containing materials such as $SrMnGe_2O_6$.[29,37] Due to the complexity of the system, constraints were applied during the refinement of the magnetic structure to achieve optimal and physically meaningful results. Best results were found when constraints caused antiparallel alignment between paired sites (Mn1$a$: Mn1$b$ and Mn2$a$: Mn2$b$) as observed in **Fig. 5.**

The refined magnetic moments for the high-pressure phase as a function of temperature are shown in **Table 1,** with the corresponding representation of the magnetic structure in **Fig. 6.** In the high-pressure monoclinic phase, the Mn sublattice changes from a puckered honeycomb lattice into a quasi-one-dimensional zigzag chain along the $b$-axis. The reduction in dimensionality likely suppresses competing exchange interactions, thereby reducing magnetic frustration and enabling magnetic order at higher temperatures. As observed within $AYbSe_2$ materials under pressure nearest neighbor exchange interactions tend to strengthen more rapidly than further neighbors, establishing a preferred hierarchy of magnetic couplings. The enhanced interaction strengths and emergence of new exchange pathways lend credence to the observed increase in the magnetic ordering temperature.

In the proposed magnetic structure, the antiparallel coupling does not occur along the nearest neighbor down the $b$-axis (2.34 Å) but instead with its next-nearest neighbor along the $ac$-plane (3.00 Å). In the high-pressure phase, the Mn atoms located along the $ac$-axis can be treated as forming dimers, analogous to those observed in $NaTiSi_2O_6$. Along these dimers, direct exchange orbital overlap stabilizes antiferromagnetic order, similar to what is observed in $CaMn_2Sb_2$.[38,39] Unlike $Ti^{3+}$, however, $Mn^{2+}$ in its high spin configuration has all five d-orbitals singularly occupied resulting in more possible exchange pathways and magnetic complexity. As a result, coupling between dimer layers remains significant, whether it be mediated through direct or superexchange interactions. The temperature range explored in the present work was limited by liquid nitrogen cooling system. Further work at lower temperatures could potentially reveal additional complexity or additional magnetic phases in CaMn$_2$Sb$_2$.

**Table 1**. Temperature dependence of Mn magnetic moment (in μ$_B$) refined at 5.9 GPa, along with total resulting moment.

| Translation | Crystal axis | 90 K | 100 K | 125 K |
|---|---|---|---|---|
| Mn1*a* | | | | |
| | a | -3.233 | -2.636 | -2.597 |
| | b | 0 | 0 | 0 |
| | c | 2.739 | 3.841 | 3.66 |
| Total moment | | 4.887 | 4.776 | 4.603 |
| Mn2*a* | | | | |
| | a | -1.550 | -1.545 | -0.917 |
| | b | -1.971 | -1.564 | -2.288 |
| | c | 4.110 | 3.223 | 3.456 |
| Total moment | | 4.509 | 3.971 | 4.286 |
| Mn1*b* | | | | |
| | a | 3.233 | 2.636 | 2.597 |
| | b | 0 | 0 | 0 |
| | c | -2.739 | -3.841 | -3.66 |
| Total moment | | 4.887 | 4.776 | 4.603 |
| Mn2*b* | | | | |
| | a | 1.550 | 1.545 | 0.917 |
| | b | 1.971 | 1.564 | 2.288 |
| | c | -4.110 | -3.223 | -3.456 |
| Total moment | | 4.509 | 3.971 | 4.286 |

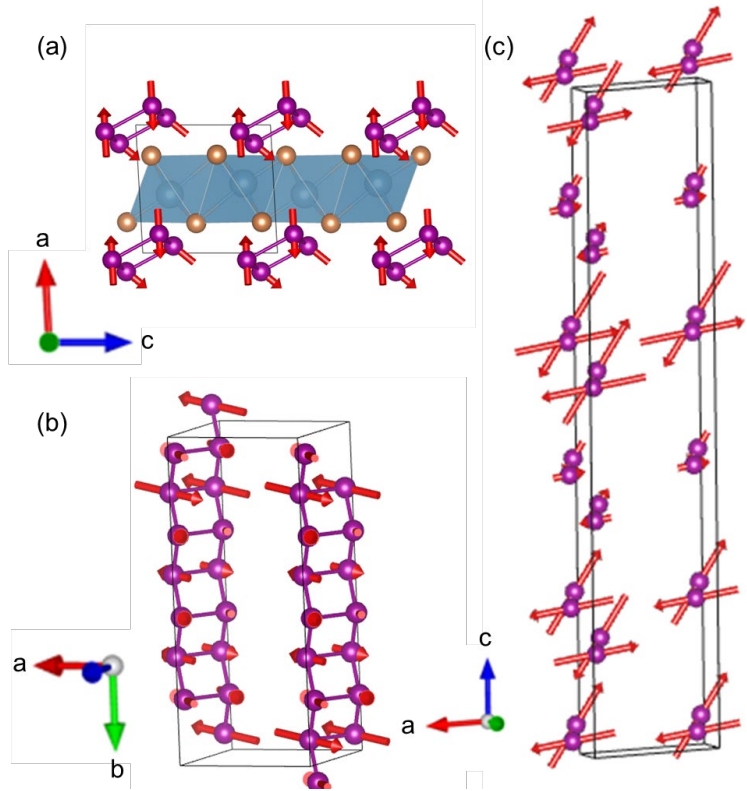

**Fig. 6** (*a*) High pressure structure with magnetic moments, cell transformed (0.5, 0, 0.5) to highlight Mn atoms. (*b*) magnetic structure showcasing zigzag chain along the *b*-axis, elongated 4-unit cells. (*c*) Magnetic structure elongated 5-unit cell in the *c*-direction.

Pressure is widely employed to tune superconductivity, either by inducing it in otherwise non-superconducting materials, enhancing it in existing superconductors, or in certain cases suppressing it entirely. The underlying mechanisms vary significantly depending on the material system. A useful comparison can be drawn between $CaMn_2Sb_2$ and $BaFe_2As_2$, the latter of which exhibits pressure-induced superconductivity.[40] Both compounds feature transition-metal atoms arranged in edge-sharing tetrahedra, separated by nonmagnetic alkaline metals; however, their structural and electronic responses to pressure diverge in key ways. In $BaFe_2As_2$, three factors are central to the emergence of superconductivity under pressure: suppression of long-range magnetic order, reduction of the Fe-Fe distance, and adjustment of the As-Fe-As bond angles toward the ideal tetrahedral value of 109.6°. At ambient pressure, $BaFe_2As_2$ undergoes a tetragonal-to-orthorhombic distortion near 135 K that stabilizes stripe-type antiferromagnetic order via Fermi surface nesting and spin-density waves. Application of pressure reduces the Fe-Fe separation and drives the As-Fe-As bond angles toward the optimal geometry, thereby destabilizing the magnetic ground state and enabling superconductivity.[40]

It is well established within the Goodenough-Kanamori-Anderson (GKA) framework that 180° superexchange geometries favor strong antiferromagnetic (AFM) interactions in predominantly spin-only systems such as high-spin $Mn^{2+}$ ($d^5$). In contrast, the structural evolution of $CaMn_2Sb_2$ under pressure suggests a very different outcome. Although the Mn-Mn distances decrease markedly, the ambient-pressure structure already possesses bond angles close to the ideal tetrahedral geometry (109.76(1)° and 109.18(1)°). Upon undergoing a volume collapse, these angles deviate strongly, shifting toward a square-pyramidal configuration with values between 97° and 105°. Moreover, a new possible Mn-Sb-Mn super-exchange pathway emerges with a bond angle of 161.3(2)°, approaching the 180° condition favored by the Goodenough-Kanamori rules. This evolution strengthens, rather than suppresses, magnetic order, which remains robust at both intermediate and high pressures. Unlike the relatively subtle symmetry changes observed in Fe-based superconductors (tetragonal-orthorhombic transitions in $BaFe_2As_2$ and $CaFe_2As_2$), $CaMn_2Sb_2$ undergoes a pronounced loss of symmetry at high pressure, further inhibiting superconductivity.[30,41,42] Computational studies of related compounds ($BaMn_2P_2$ and $BaMn_2As_2$) predict that their antiferromagnetic states persist up to at least 127 GPa, with superconductivity only possible at extreme conditions.[43] Collectively, these results indicate that the magnetic moments in $CaMn_2Sb_2$ are more localized and stabilized by super-exchange interactions rather than Fermi surface nesting, rendering the antiferromagnetic order highly resilient to pressure and precluding the emergence of superconductivity.

**Conclusion**

In summary, $CaMn_2Sb_2$ exhibits a rich interplay between structural, electronic, and magnetic degrees of freedom under applied pressure. Single-crystal X-ray diffraction measurements reveal a pressure-induced first-order phase transition from a layered trigonal structure (space group *P-3m*1) to a monoclinic structure (space group *P2$_1$/m*) at approximately 5.4 GPa, accompanied by a ~7% volume collapse and substantial crystallographic distortion. Residual electron density analysis indicates the emergence of charge instabilities near the transition point, manifested as anisotropic charge localization along Mn-Sb linear chains. Detailed bonding analysis highlights pressure-induced reconfiguration of Mn-Sb orbital interactions, driving the system toward a distorted square-pyramidal geometry. High-pressure neutron diffraction measurements further confirm the structural phase transition and uncover the emergence of a one-dimensional incommensurate magnetic order above the transition pressure. The transition from a Néel-type antiferromagnetic order at ambient conditions to a quasi-commensurate spin density wave with a complex propagation vector illustrates the sensitivity of magnetic ground states to lattice and bonding distortions. Importantly, the high-pressure monoclinic phase hosts zigzag Mn chains and features antiferromagnetic coupling along the *ac*-plane, likely facilitated by enhanced Mn-Mn orbital overlap and symmetry-equivalent coordination environments. These findings provide critical insights into the mechanisms by which antiferromagnetic insulators accommodate electronic and structural instabilities under pressure, and they underscore the role of lattice geometry and bonding topology in dictating the emergence of nontrivial magnetic phases in Mn-based layered pnictides.

## Supporting Information

This material is available free of charge.

Crystal data and structure refinement of $CaMn_2Sb_2$ at various pressures; Atomic coordinates and isotropic atomic displacement parameters ($Å^2$); Resistivity measurement on synthesized crystal of $CaMn_2Sb_2$; Lattice parameters as a function of pressure; Lowest occupied molecular orbitals (LOMOs) of Mn@Sb4 layers.

## Acknowledgement

The work at Michigan State University was supported by the U.S.DOE-BES under Contract DE-SC0023648. S.-Y. X. was supported by the NSF Career DMR-2143177. A portion of this research used resources at the Spallation Neutron Source, a DOE Office of Science User Facility operated by the Oak Ridge National Laboratory. The beam time was allocated to the SNAP diffractometer under proposal number IPTS-33086. This material is based upon work supported by the U.S. Department of Energy, Office of Science, Office of Workforce Development for Teachers and Scientists, Office of Science Graduate Student Research (SCGSR) program. The SCGSR program is administered by the Oak Ridge Institute for Science and Education (ORISE) for the DOE.

## Conflict of Interest

The authors declare no conflict of interest.

**TOC graphic**

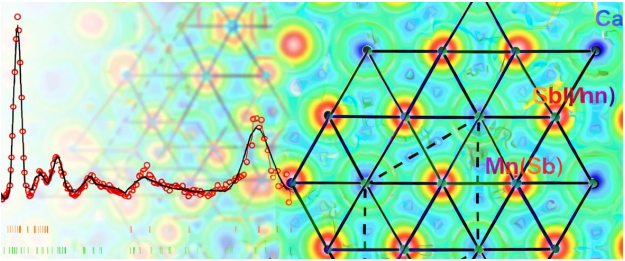

# Supplementary Information

# Pressure-Induced Chemical Bonding Effects on Lattice and Magnetic Instabilities in Antiferromagnetic Insulating CaMn$_2$Sb$_2$

Matt Boswell[1], Antonio M. dos Santos[2], Mingyu Xu[1], Madalynn Marshall[3], Su-Yang Xu[4], Weiwei Xie[1*]

5. Department of Chemistry, Michigan State University, East Lansing, MI 48824 USA
6. Neutron Scattering Division, Oak Ridge National Laboratory, Oak Ridge, TN 37831 USA
7. Department of Chemistry and Biochemistry, Kennesaw State University, Kennesaw, GA 30144 USA
8. Department of Chemistry and Chemical Biology, Harvard University, Cambridge, MA 02138, USA

Corresponding Author: Weiwei Xie (xieweiwe@msu.edu)

**Table of Contents**

Table S1. The crystal structure and refinement of CaMn$_2$Sb$_2$ at room temperature and various pressures. Values in parentheses are estimated standard deviation from refinement.

| Pressure | 0 GPa | 4.5 GPa | 5.9 GPa |
|---|---|---|---|
| Space Group | P-3m1 | P-3m1 | P2$_1$/m |
| Lattice Parameters (Å) | a = 4.5359(1)<br>c = 7.4930(3) | a = 4.4097(3)<br>c = 7.278(1) | a = 7.2708(15)<br>b = 4.1436(8)<br>c = 7.3702(15)<br>β = 93.88(3)° |
| Volume (Å$^3$) | 133.51(1) | 122.57(3) | 221.54(8) |
| Absorption coefficient | 15.407 mm$^{-1}$ | 16.783 mm$^{-1}$ | 9.285 mm$^{-1}$ |
| F (000) | 172 | 172 | 172 |
| θ range (°) | 2.72 to 41.14 | 5.34 to 36.26 | 5.55 to 36.23° |
| Reflections collected | 8576 | 3285 | 1394 |
| Independent reflections | 382 | 152 | 293 |
| Refinement method | F$^2$ | F$^2$ | F$^2$ |
| Data/restraints/parameters | 382/0/9 | 152/0/16 | 293/0/17 |
| Final $R$ indices | $R_{1(I>2\sigma(I))}$ = 0.0238;<br>$wR_{2(I>2\sigma(I))}$= 0.0558<br>$R_{1(all)}$ = 0.0257;<br>$wR_{2(all)}$ = 0.0562 | $R_{1(I>2\sigma(I))}$ = 0.0651;<br>$wR_{2(I>2\sigma(I))}$= 0.1585<br>$R_{1(all)}$ = 0.1377;<br>$wR_{2(all)}$ = 0.2416 | $R_{1(I>2\sigma(I))}$ = 0.4012;<br>$wR_{2(I>2\sigma(I))}$= 0.6405<br>$R_{1(all)}$ = 0.6129;<br>$wR_{2(all)}$ = 0.7301 |
| Largest diff. peak &hole | +5.245 e$^-$/Å$^3$<br>-0.719 e$^-$/Å$^3$ | +7.869 e$^-$/Å$^3$<br>-10.165 e$^-$/Å$^3$ | +1.861 e$^-$/Å$^3$<br>-3.173 e$^-$/Å$^3$ |
| R.M.S. deviation | 0.318 e$^-$/Å$^3$ | 0.947 e$^-$/Å$^3$ | 1.028 e$^-$/Å$^3$ |
| Goodness-of-fit on F$^2$ | 1.087 | 1.264 | 2.529 |

**Table S2.** Atomic coordinates and equivalent isotropic atomic displacement parameters (Å$^2$) of CaMn$_2$Sb$_2$ at room temperature and various pressures. ($U_{eq}$ is defined as one-third of the trace of the orthogonalized $U_{ij}$ tensor.)

**0 GPa**

| Atoms | Wyck. | x | y | z | Occ. | $U_{eq}$ |
|---|---|---|---|---|---|---|
| Ca | 1a | 0 | 0 | 0 | 1 | 0.015(1) |
| Mn | 2d | 1/3 | 2/3 | 0.6230(1) | 1 | 0.015(1) |
| Sb | 2d | 1/3 | 2/3 | 0.2514(1) | 1 | 0.013(1) |

**4.5 GPa**

| Atoms | Wyck. | x | y | z | Occ. | $U_{eq}$ |
|---|---|---|---|---|---|---|
| Ca  | 1a | 0   | 0   | 0         | 1     | 0.035(6)  |
| Mn1 | 2d | 1/3 | 2/3 | 0.4027(2) | 0.813 | 0.001(1)  |
| Mn2 | 2d | 1/3 | 2/3 | 0.280(5)  | 0.132 | -0.109(1) |
| Mn3 | 2d | 1/3 | 2/3 | 0.078(7)  | 0.067 | -0.103(1) |
| Sb  | 2d | 1/3 | 2/3 | 0.7512(8) | 1     | 0.040(2)  |

**5.9 GPa**

| Atoms | Wyck. | x | y | z | Occ. | $U_{iso}$ |
|---|---|---|---|---|---|---|
| Ca  | 2e | 0.028(1) | ¼ | 0.710(1) | 1 | 0.09(1) |
| Mn1 | 2e | 0.535(1) | ¼ | 0.797(1) | 1 | 0.20(1) |
| Mn2 | 2e | 0.420(1) | ¼ | 0.442(1) | 1 | 0.14(1) |
| Sb1 | 2e | 0.273(1) | ¼ | 0.074(1) | 1 | 0.16(1) |
| Sb2 | 2e | 0.813(1) | ¼ | 0.364(1) | 1 | 0.17(1) |

**Figure S1.** Resistivity measurement on synthesized crystal of CaMn₂Sb₂

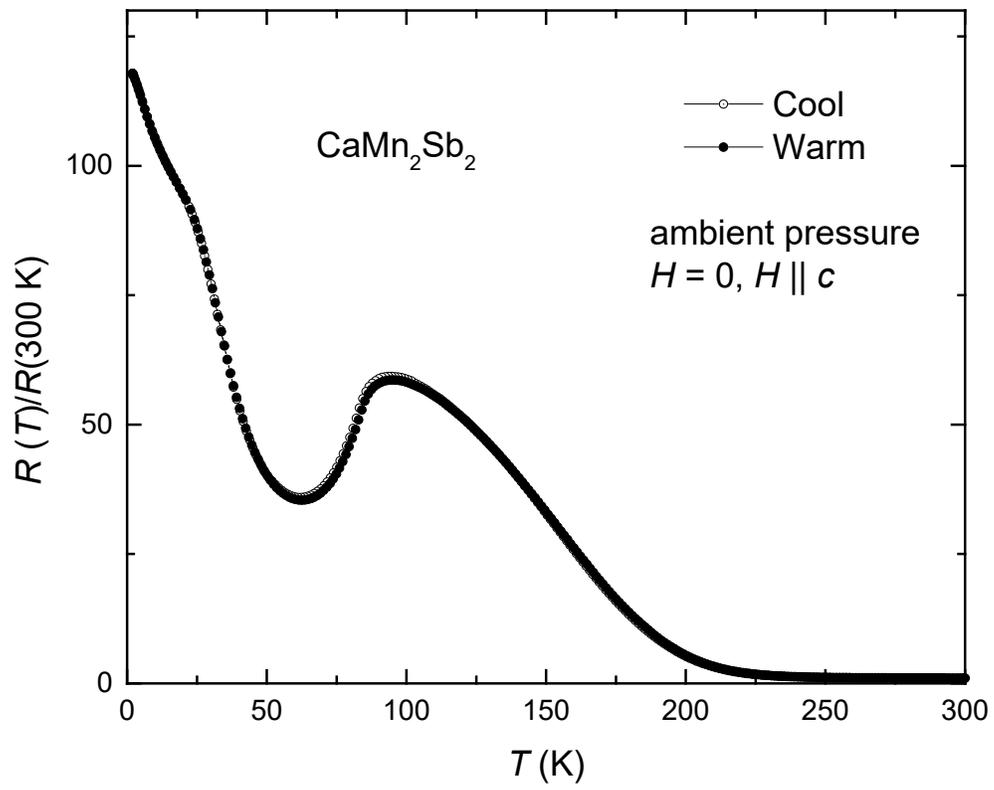

**Figure S2.** Normalized lattice parameters to initial lattice values at 0 GPa.

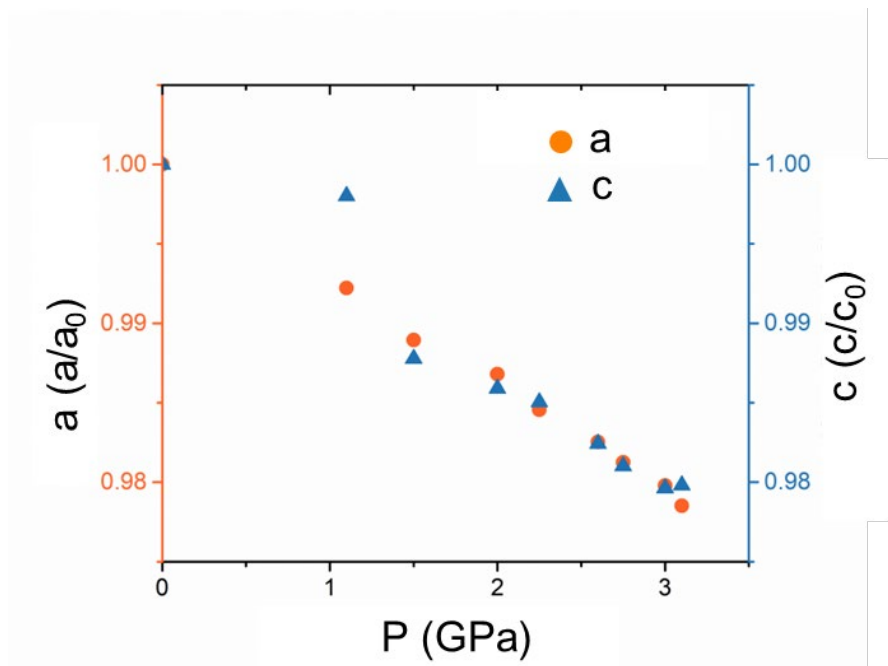

**Figure S3.** Lowest occupied molecular orbitals (LOMOs) of Mn@Sb4 layers in (*a*) the tetragonal phase and (*b*) the trigonal phase.

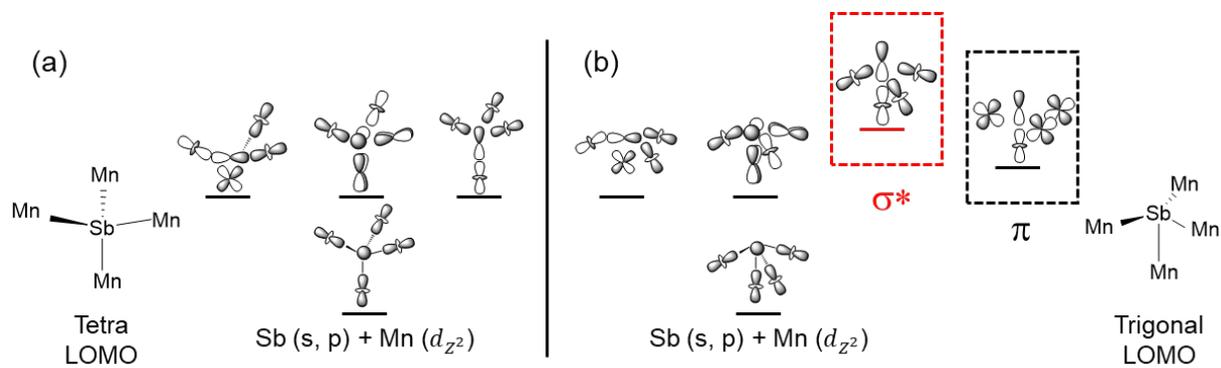